\documentclass[a4paper]{jpconf} % letterpaper could be used instead of a4paper
\usepackage{graphicx}
\begin{document}
%You can use \jpcs to insert 'Journal of Physics: Conference Series' in italics
\title{Study and perspective on neutron beam divergence improvement achievable by the combination of two or more neutron collimating systems}

\author{Oriol Sans-Planell$^{1,2}$, Francesco Cantini$^{3,4}$, Marco Costa$^{1,2}$, Francesco Grazzi$^{3,4}$, Manuel Morgano$^{5}$, Masako Yamada$^{6}$}

\address{${}^{1}$Università degli Studi di Torino, Via Pietro Giuria 1, 10125 Torino, Italy\\${}^{2}$INFN Sezione di Torino, Via Pietro Giuria 1, 10125 Torino, Italy\\${}^{3}$Consiglio Nazionale delle Ricerche, Istituto dei Sistemi Complessi, Via Madonna del Piano 10, 50019 Sesto Fiorentino, Italy\\${}^{4}$INFN Sezione di Firenze, Via Madonna del Piano 10, 50019 Sesto Fiorentino, Italy\\${}^{5}$European Spallation Source, Partikelgatan 2, 22484 Lund, Sweden\\${}^{6}$Paul Scherrer Institute, Forschungsstrasse 111, 5232 Villigen, Switzerland}

\ead{oriol.sansplanell@to.infn.it}

\begin{abstract}%you must include an abstract
This communication presents the results obtained at an experimental campaign at PSI BOA beamline using the combination of the ANET Compact Neutron Collimator (CNC) with the actual BOA pin-hole system. Through extensive resolution campaigns, it has been possible to quantify and understand the effects of improvement on the beam divergence when combining the two collimating systems. A new theoretical approach to this problem is described and discussed.
The effect is expected not to be limited to the specific case that has been studied at PSI BOA but to have a more general validity for neutron collimation systems.
\end{abstract}

\section{Introduction}
The study, here reported, derives its motivation from measurements performed at the TU-Delft FISH beam-line \cite{fish}, where an unexpected improvement on the quality of the beam collimation \cite{anet1}, beyond what a simple geometrical model could predict, have been shown. This indicates that more complicated effects arise when combining two collimation systems. \newline
Commonly a geometrical model states that, when two collimators are concatenated in a set-up, the one with the higher L/D will determine the final collimation power of the system. 
This assumption has certainly some fundamental of truth, but our measurements showed also its limitations, since the model only takes into consideration the geometrical properties of the system but it does not include that the divergence distribution of the neutrons after the first collimator is different from that of the original source. 
In order to replicate the results obtained at TU-Delft and a better understanding of the observed behavior, a more extensive measurement campaign has been done at the BOA beam-line \cite{boa}, at the Paul Scherrer Institut. \\
A hypothesis of the formal description of the convolutional effect between both collimators is described in the following sections.\\

\section{The experiment}
The experiment has been performed at the Paul Scherrer Institut, at the BOA facility \cite{boa}. It is a multi-purpose beamline composed by a 18$m$ long tube with a cold neutron flux of $10^8cm^{-2}s^{-1}mA^{-1}$ at the exit window. The beamline is composed by a series of neutrons guides and a bender, which creates some alterations on the beam divergence. The beam exit window can also be confined, through a pinhole, to different sizes, up to 40 x 150 $mm^2$. For the measurements, the beam was restricted to a rectangular shape of 40 x 80 $mm^2$.\newline
\begin{figure}[h]
    \centering
    \includegraphics[width=0.9\textwidth]{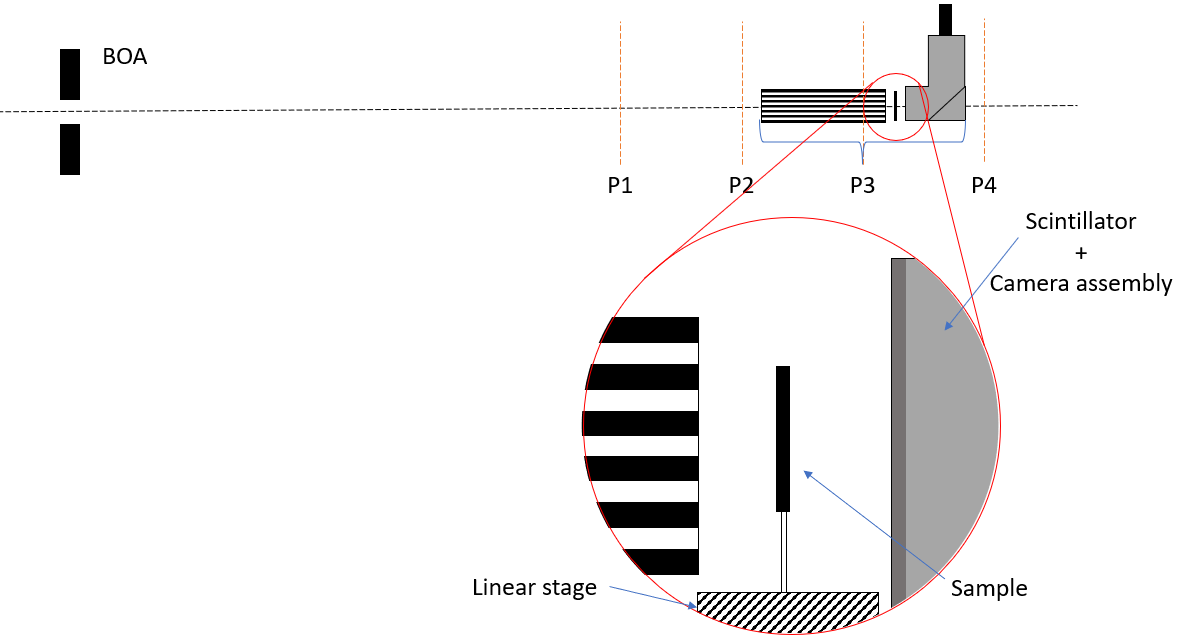}
	\caption{Design of the experimental set-up present at the campaign.}
	\label{fig:BOA-setup}
\end{figure}
The set-up used was analogous to the one already defined on previous experimental campaigns \cite{anet1,pavia}. It made use of three towers - labeled as T4, T5 and T6 -. Each one has a Z-axis moving stage parallel to the beam: the first one (T4) held the Stewart Platform and the collimator, the second one (T5) held a secondary X-axis linear stage with the samples, and the third one (T6) held the detector. The three T stages were configured so that the distance between the exit window of the collimator and the scintillator of the detector was 190$mm$. The two standard reference samples were chosen to be a gadolinium knife edge and a Siemens star, both manufactured at the PSI \cite{samples}, and mounted in the middle stage (T5). The stage moved the samples at 8 different positions, from 10$mm$ up to 80$mm$ from the scintillator at 10$mm$ steps - labeled as micro-positions -. The T5 X-axis motor served to change from one sample to the other.
\begin{table}[h]
\centering
\begin{tabular}{c|cccl}
Position & T4   & T5   & T6   &  \\ \cline{1-4}
1        & 4480 & 4920 & 5360 &  \\
2        & 5480 & 5920 & 6360 &  \\
3        & 6480 & 6920 & 7360 &  \\
4        & 7480 & 7920 & 8360 & 
\end{tabular}
\caption{Distance from the primary pin-hole to each motor (in $mm$) at each macro-position (1-4) for the measurements at BOA.}
\label{tbl:motor-positions}
\end{table}
Table \ref{tbl:motor-positions} describes the position of each stage, from T4 to T6 at each macro-position with respect to the primary pin-hole. \newline
The camera used to acquire the images was an ANDOR 16-bit cooled CCD with a field of view of 4.00 x 4.00 $cm^2$ over a resolution of 1024 x 1024, leading to a projected pixel size at the scintillator of 33.65$\mu m$. The BOA system was controlled through the software package NIKOS, which is a custom distribution of the open source EPICS \cite{epics}, while the Stewart platform was controlled via LAN using Labview.

\section{Experimental results}
\begin{figure}[h]
    \centering
    \includegraphics[width=0.9\textwidth]{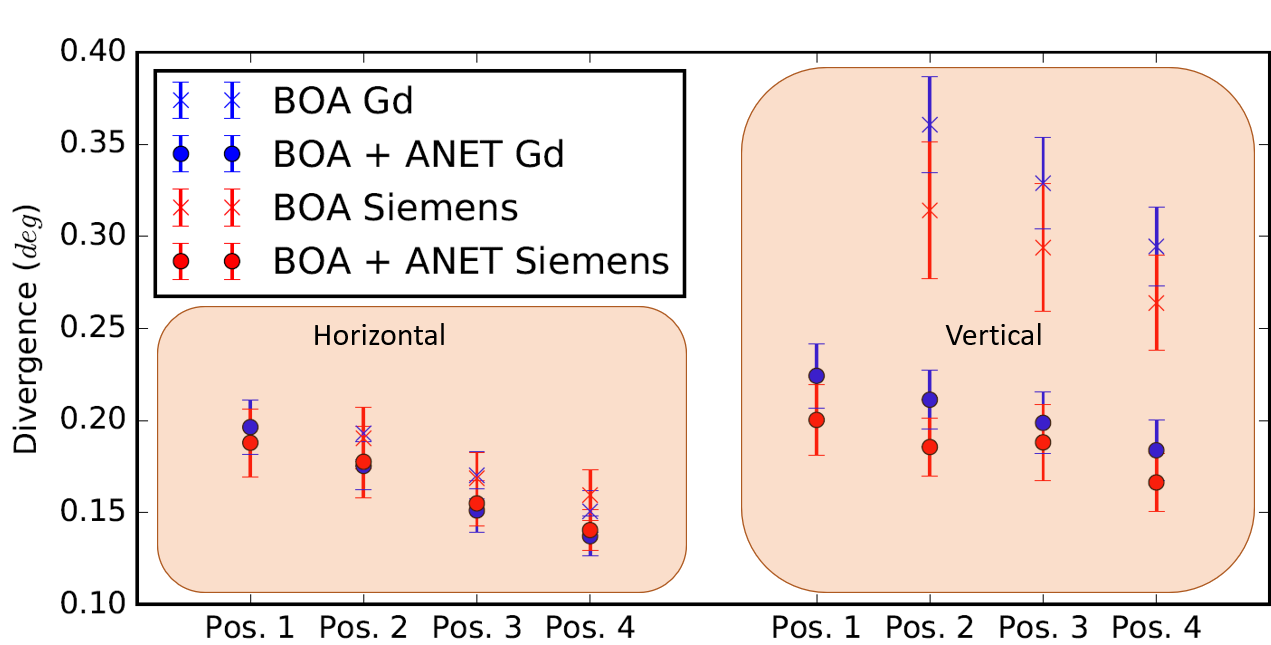}
	\caption{Graph showing the divergence angle calculated on each macro-position with both the Siemens star and the gadolinium knife-edge.}
	\label{fig:BOA-recap}
\end{figure}

The measurements taken with both Siemens Star and gadolinium knife-edge reference samples have been analysed following the procedure in \cite{anet1}, and the results are shown in figure \ref{fig:BOA-recap}
The expected beam divergence should not be affected by a collimator whose collimation power is smaller than the beam divergence itself. The measurements shown in figure \ref{fig:BOA-recap} point at different conclusions. Not only the divergence is improved when the ANET CNC is included within the beamline, but it greatly reduces the beam angular divergence beyond its expected capabilities. A second appreciation is that the absolute improvement is not the same for the vertical and horizontal axis, despite the collimator applying the same potential correction factor by having a square section geometry. A hypothesis why this phenomenon happens will be studied in the following section.

\section{Mathematical description}
\begin{figure}[htb]
    \centering
    \includegraphics[width=0.7\textwidth]{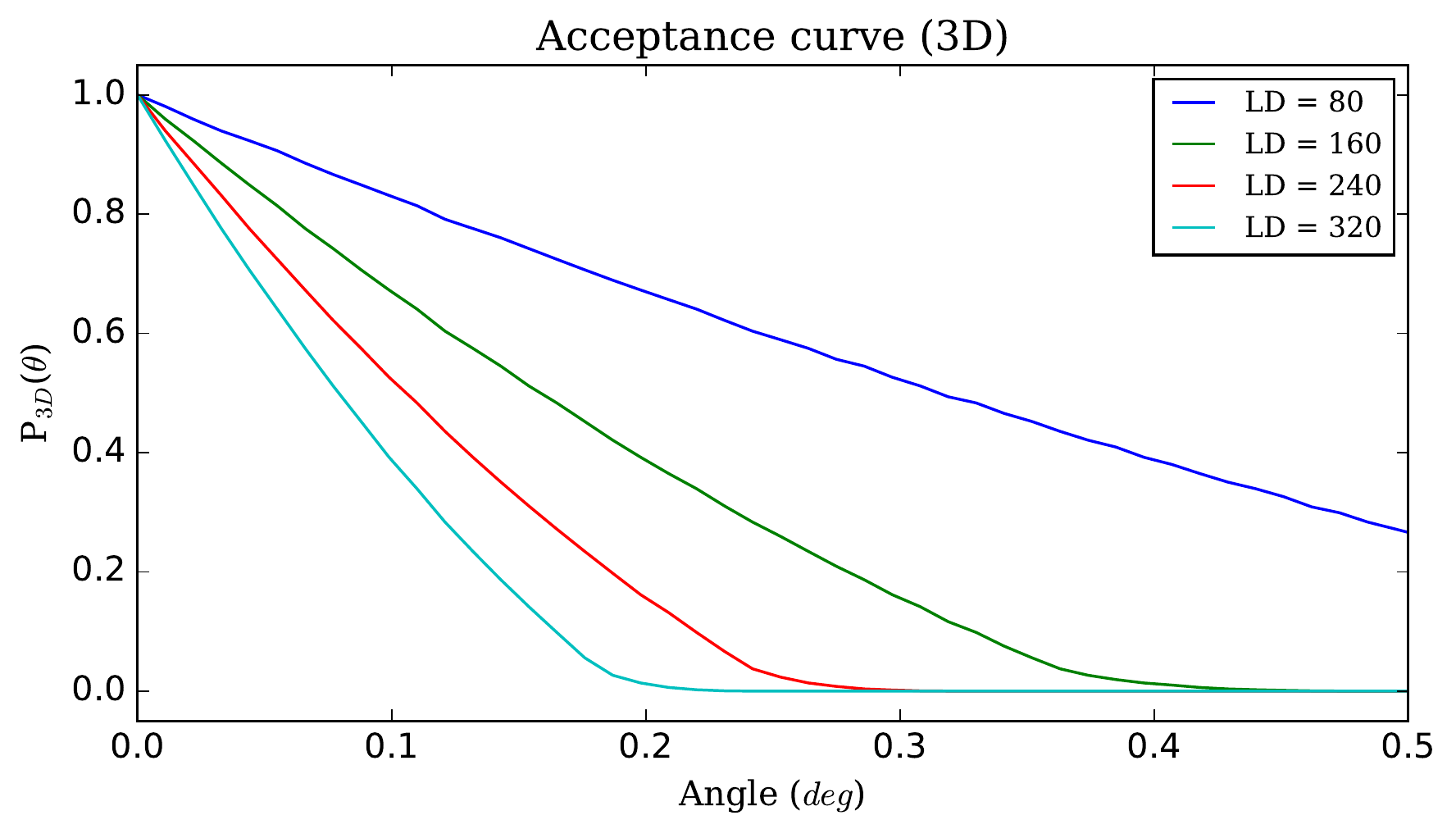}
	\caption{Different acceptance curves calculated in 3D while varying the L/D factor. The probability of survival of the neutrons decrease with the increasing angle with respect to the propagation axis.}
	\label{fig:acceptance-3d}
\end{figure}

The approximate geometrical behaviour that a neutron would have when passing through a collimator made of a perfect absorbing material can be described through the following formula, for the 2D case: 
\begin{equation}
    P_{2D}(\theta)=\int_{x} \rho(x,\theta) \delta(\theta|x)dx
\end{equation}
This equation describes the probability distribution of a neutron to survive the collimator, $P(\theta)$, calculated through the integral of the initial space and angular distribution, $\rho(x,\theta)$, and a step function which can either take values 0 or 1, depending on whether the neutrons are absorbed or not, which is determined by the angle given a certain position. To calculate the shape of this function, a very simple empirical simulation has been designed: since the probability of passing through the collimator is either 0 or 1, a simple channel was designed, with given characteristics (L and D), and every possible neutron within an angular range of [-$\theta$,+$\theta$] and an spatial range of [$-\frac{D}{2}$,$+\frac{D}{2}$) was simulated. The output of this model is the distribution $P(\theta)$ which describes the probability of survival through the given collimator. 
\begin{figure}[h]
    \centering
    \includegraphics[width=0.7\textwidth]{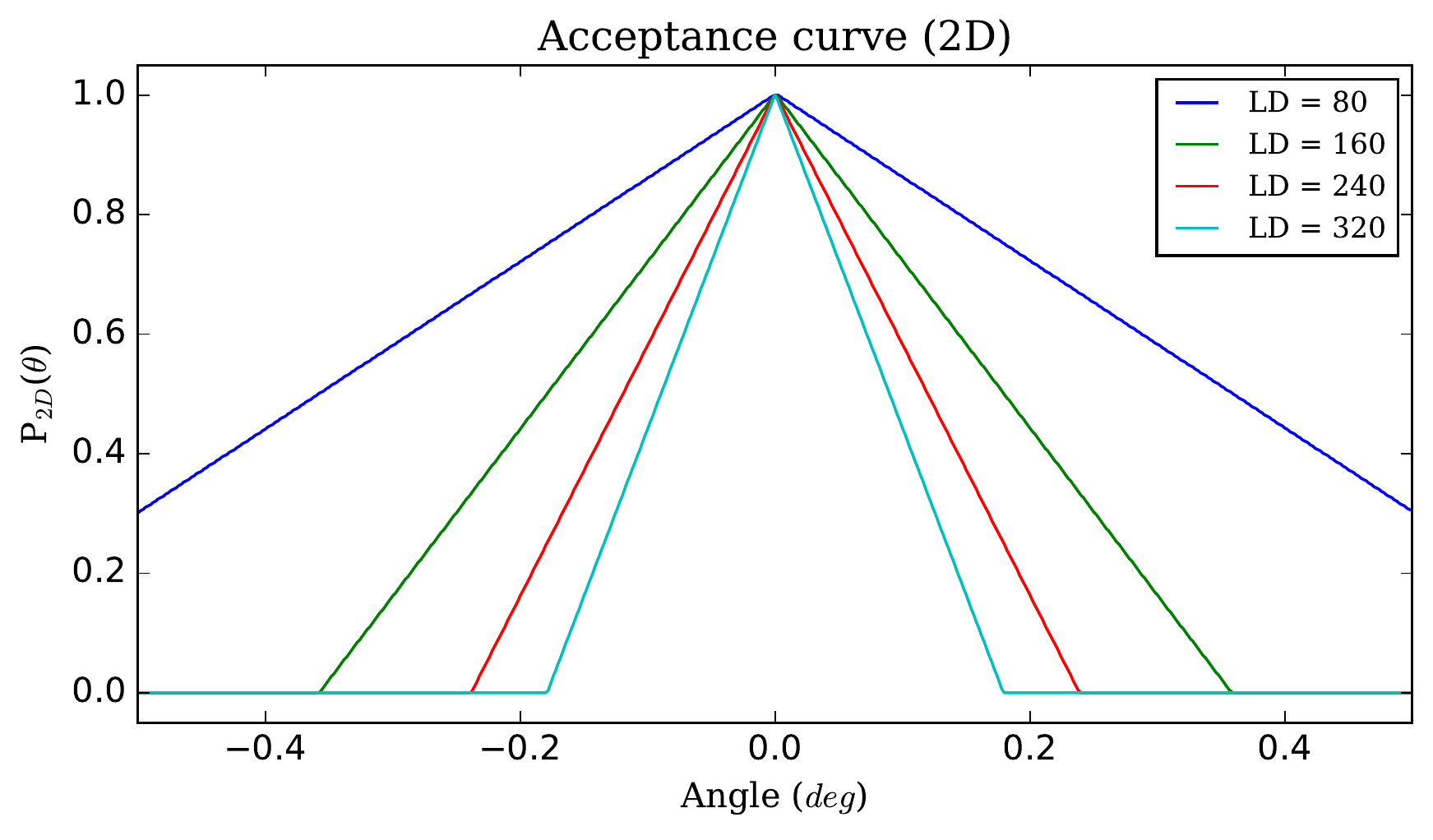}
	\caption{2D acceptance curves calculated at different values of L/D. The graphic shows the probability of "survival" of a neutron depending on the divergence angle.}
	\label{fig:acceptance-2d}
\end{figure}
Figure \ref{fig:acceptance-2d} shows some acceptance curves generated with different L/D values. These curves can be parametrised with a straight line function, which is: 
\begin{equation}
    P_{2D}(\theta)=\theta\Big(arctan(\frac{1}{L/D})\Big)^{-1}+1
\end{equation}
The value $\Big(arctan(\frac{1}{L/D})\Big)^{-1}$ essentially describes the limit angle over which the probability of survival is 0\footnote{As it is a probability, negative values are equal to 0.}.\newline
The model can be expanded to 3D, in order to evaluate the total divergence, instead of evaluating only one axis. When treating the collimator in 3D, two more variables are added: there are two spatial coordinates $(x,y)$ and two angular coordinates, the divergence from the propagation axis, $\theta$ and the rotation along the propagation axis, $\phi$, which is sampled in $[0,2\pi]$. The formula that describes the mathematical model is:
\begin{equation}
    P_{3D}(\theta)=\int_{x}\int_{y}\int_{\phi} \rho(x,y,\theta,\phi) \delta(\theta,\phi|x,y) dx dy d\phi
    \label{eq:acceptance-2d}
\end{equation}
The solution of $P_{3D}(\theta)$, contrary to the 2D version, depends heavily on the geometry of the collimator. In the ANET CNC, the 3D distribution follows: 
\begin{equation}
    P_{3D}\Big(\sqrt{\theta_{1}^{2}+\theta_{2}^{2}}\Big) = P_{2D}(\theta_{1})\cdot P_{2D}(\theta_{2})
    \label{eq:acceptance-3d}
\end{equation}
This relation is due to the square geometry of the collimator individual channels. Figure \ref{fig:acceptance-3d} has the result of some calculations of the 3D acceptance curves. 
Contrary to the 2D case, in the 3D version the $\theta$ value has a minimum value of 0. This is due to the $\phi$ being sampled from 0 to $2\pi$. \newline
This function acts as a mask over the incident beam, altering the shape regardless of the initial beam divergence. To test this hypothesis, the angular distributions from the BOA facility simulation, done in McStas, have been used. The beam divergence distributions before and after the collimator have been simulated. If the mathematical formula is correct, multiplying the angular distribution present at the entrance window of the collimator by the mask should result in the beam's angular distribution at the exit window. This is discussed in the following section.

\section{Testing the Acceptance Curve parametrisation}
The set-up has been simulated using McStas \cite{mcstas} in 2 positions: with the CNC at 5.23$m$ from the BOA exit window, and at 20$cm$. The calculation test has been done considering both the 2D and 3D cases.

\subsection{The 2D cases}

\begin{figure}[h]
    \centering
    \includegraphics[width=1\textwidth]{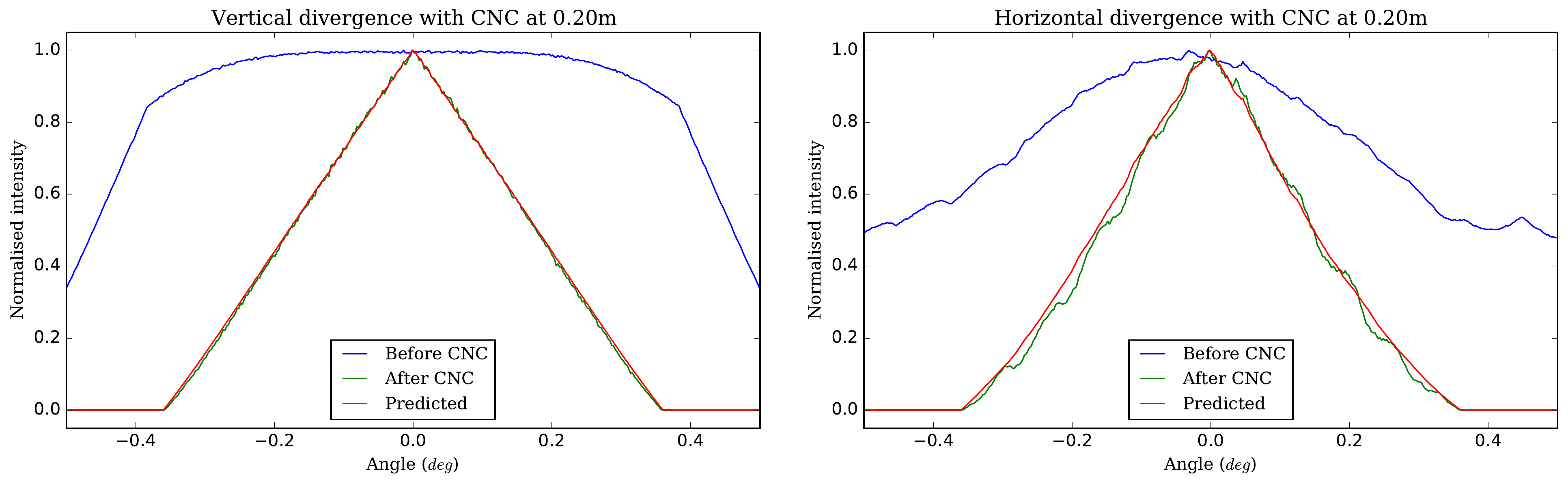}
	\caption{Vertical and horizontal angular divergence distribution plots before and after the ANET CNC along with the predicted distribution. Simulation done with the CNC at 20 $cm$ from the exit window of BOA.}
	\label{fig:acceptance-test1}
\end{figure}

\begin{figure}[h]
    \centering
    \includegraphics[width=1\textwidth]{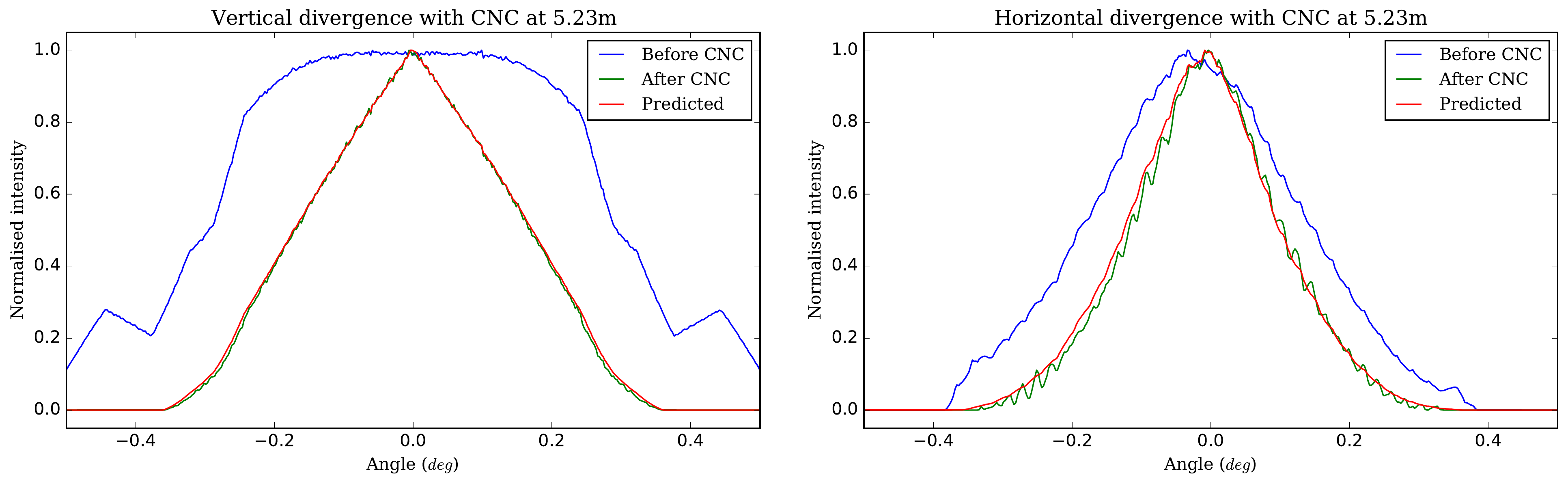}
	\caption{Vertical and horizontal angular divergence distribution plots before and after the ANET CNC along with the predicted distribution. Simulation done with the CNC at 5.23 $m$ from the exit window of BOA.}
	\label{fig:acceptance-test2}
\end{figure}

Both close (figure \ref{fig:acceptance-test1}) and far (figure \ref{fig:acceptance-test2}) configurations have been simulated. The plots in the figures have been normalised to the maximum value of each plot, in order to compare the shapes of the distributions. All cases show a very good agreement between the calculated distributions and the simulated ones. Those plots render evident the difference in the BOA facility between the vertical and horizontal distributions. The bender present along the guide creates distortions on the horizontal axis, as expected. The plots of the vertical divergence shows a wide and uniform distribution before the CNC, which is strongly reduced by the collimator down to the shape of the acceptance curve. On the horizontal axis, instead, the irregular shape of the angular distribution is still very well predicted by the model, as it adapts onto the shape of the acceptance curve. 
When the ANET CNC is installed far from the exit window of BOA (figure \ref{fig:acceptance-test2}), the pin-hole collimation of the beam reduces the maximum angular divergence to less than 0.5 degrees. On the vertical axis, the distribution has, at the tails, two secondary peaks, artifacts generated by the beam construction. Those artifacts get eliminated by the collimator, which also shrinks the distribution.
On the horizontal axis, the effect of collimation is less prominent than on the vertical axis, but there is still a reduction of the more divergent neutrons.

\subsection{The 3D cases}
\begin{figure}[h]
    \centering
    \includegraphics[width=1\textwidth]{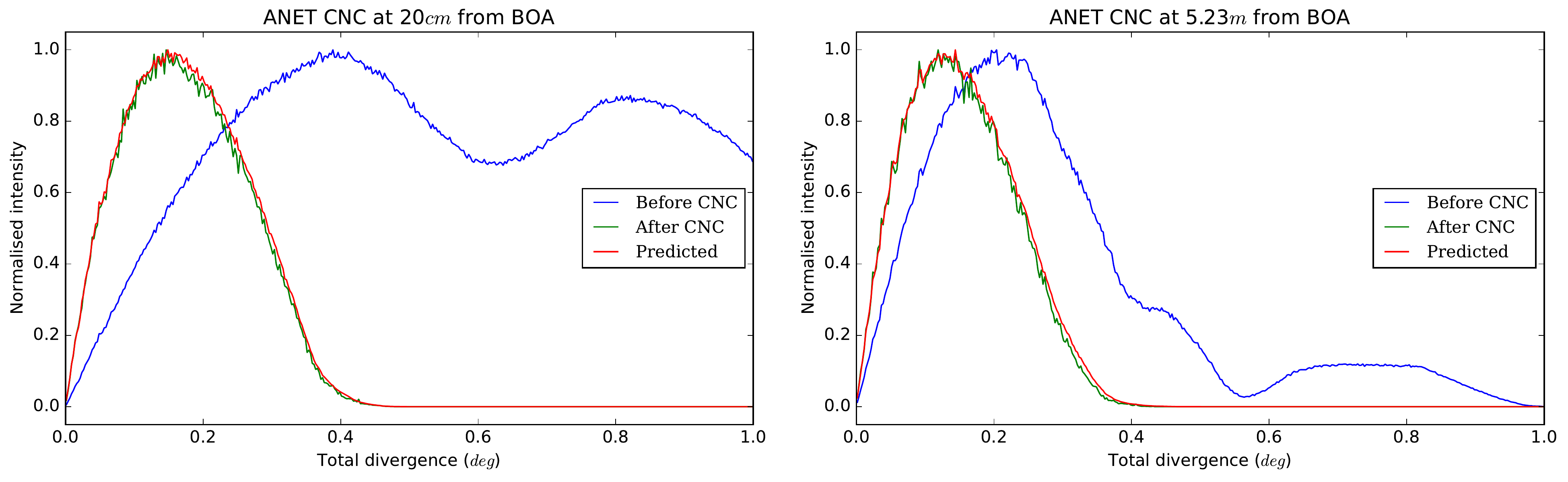}
	\caption{Results of the simulations of the 3D neutron angular distribution before and after the ANET CNC, along with the analytical calculation of the resulting angular distribution.}
	\label{fig:acceptance-test-3d}
\end{figure}

\begin{figure}[h]
    \centering
    \includegraphics[width=0.7\textwidth]{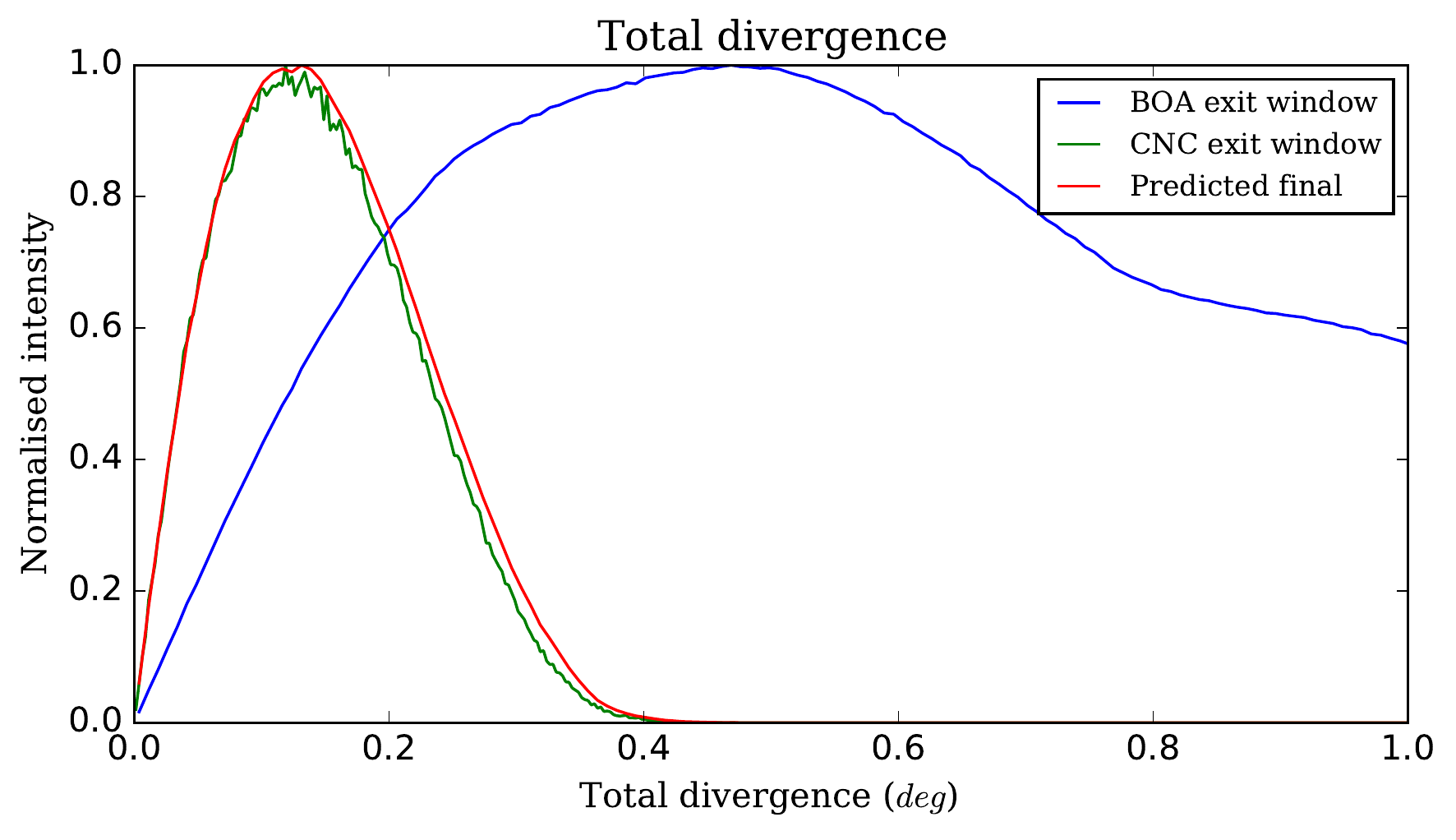}
	\caption{Results of the simulations of the 3D neutron angular distribution at the BOA exit window and after the ANET CNC. The graph also shows the calculated neutron distribution after the ANET CNC.}
	\label{fig:acceptance-double}
\end{figure}

Regarding the 3D calculation, the formula used to extract the analytical shape is reported in equation \ref{eq:acceptance-3d}, leading to the following results in figure \ref{fig:acceptance-test-3d}. The effect of the ANET CNC is very apparent in both cases, but specially when the collimator is close to the exit window of BOA. All the neutrons beyond the 0.50 limit angle are absorbed, while the whole distribution shifts towards smaller divergence angles. On the far set-up configuration, the angular distribution without the collimator is more irregular, and it becomes smoother when the ANET CNC is present in the set-up. In both close and far configurations, the model accurately predicts the impact of the ANET CNC on the angular divergence of the beam.\\
The same test can be applied when calculating the neutron angular distribution while combining two curves of acceptance: the pin-hole effect from the BOA exit window to the ANET CNC, and the ANET CNC itself. The result is shown in figure \ref{fig:acceptance-double}.
The normalised calculated curve adapts fairly well to the simulated angular distribution obtained from the McStas simulations.

\section{Conclusions}
A first important consequence of the curve of the mathematical acceptance model is the fact that the collimator has an active effect in altering the quality of the image, regardless of the angular divergence of the initial beam. A second consequence is that the collimator doesn't alter the absolute divergence if the beam is more collimated, as the endpoint of the distribution after the collimator doesn't change. The improvement in the image quality is due to the statistical reduction of the number of neutrons present at high divergence angle, not because the collimator fully eliminates them. \newline
This hypothesis is not limited to the ANET CNC, as at no point within the calculations the full ANET geometry has been imposed. The model has been done for a single collimation channel, and the calculations  are perfectly applicable to a circular collimator or a pinhole. Using this hypothesis, the combination of more than one pinhole, or with the ANET collimator itself, can be used as a beam divergence corrector. Equations \ref{eq:acceptance-2d} and \ref{eq:acceptance-3d} imply that the effect of the collimator is null on the perfectly straight neutrons, while the greater the angle, the better the absorption, up to the value of 1 (perfect absorption) when the angle is beyond the geometrical acceptance of the collimator. The effect that the collimation has on the neutron angular distribution implies that, at some point, the most-divergent neutrons will be "smoothed out" and they will be not statistically significant to contribute above the image noise, becoming than not distinguishable. The macroscopic consequence of this effect is that the measured  neutron divergence is better than what can be calculated.
\ack%this is an unnumbered acknowledgement section
The authors would like to acknowledge the INFN for the financial support of the ANET project. A special thanks also to the support given by members of the Paul Scherrer Institute, for their insights and help on the experimental campaigns, in particular Pierre Boillat, Pavel Trtik, Anders Kaestner and Uwe Filges.

\end{document}